\documentclass[12pt]{article}
\usepackage{epsfig}
\begin{document}
\title{On the nature of fully-charmed four-quark exotic state $X(6900)$ from its photoproduction off nuclei}
\author{E. Ya. Paryev\\
{\it Institute for Nuclear Research of the Russian Academy of Sciences}\\
{\it Moscow, Russia}}

\renewcommand{\today}{}
\maketitle

\begin{abstract}
The possibility to study the nature of fully-charmed tetraquark state $X(6900)$ from its inclusive photoproduction off nuclei near the kinematic threshold is investigated within the collision model based on the nuclear spectral function. The model accounts for $X(6900)$ production in direct photon--nucleon interactions as well as three different scenarios for its internal structure: compact tetraquark, molecule of the two hidden-charm $J/\psi$ and $\psi(3770)$ mesons and the mixture of both of them. We calculate within these scenarios the absolute and relative excitation functions on $^{12}$C and $^{184}$W nuclei at near-threshold photon energies of 30--40 GeV, the absolute momentum differential cross sections and ratios of them for the $X(6900)$ meson production off these target nuclei at laboratory polar angles of 0$^{\circ}$--5$^{\circ}$ and for photon energy of 35 GeV as well as the A-dependences of its transparency ratios at photon energy of 35 GeV. We show that these observables reveal a definite sensitivity to the $X(6900)$ intrinsic structure. Therefore, they might be useful for the determination of this structure from the comparison of them with the experimental data from the future experiments at the upcoming experimental facilities, such as the planned electron-ion colliders in the United States and China.
\end{abstract}

\newpage

\section*{1. Preamble}

\hspace{1.5cm} Since the discovery of the $X(3872)$ resonance at Belle/KEK in
2003 [1] as a narrow peak in the vicinity of the $D^0{\bar D}^{*0}$ (${\bar D}^0{D}^{*0}$) mass threshold in the ${J/\psi}{\pi^+}\pi^-$ invariant mass distribution in exclusive $B^{\pm} \to K^{\pm}X(3872) \to K^{\pm}({J/\psi}{\pi^+}\pi^-)$ decays, many exotic hadrons, which cannot be explained by the
conventional quark model for quark-antiquark mesons and three-quark baryons, have been observed in high-energy
experiments. They are composed of four or five quarks (and antiquarks) and usually are named as tetraquark and pentaquark states, respectively. Thus, many unconventional states involving charm or bottom quarks, charged or neutral, the so-called $X, Y, Z$ mesons as well as the hidden-charm non-strange $P_c$ and strange $P_{cs}$ pentaquark states,
the doubly-charmed tetraquark $T^+_{cc}(3875)$ state and so on have been observed in various experiments,
as summarized in recent reviews [2--5]. They are an important topic in hadron physics by now since
the studying of their properties is very essential for understanding the perturbative and nonperturbative aspects of QCD. The discovery of these exotic QCD states has motivated an extensive theoretical efforts to understand their properties [6--12]. But, in spite of these efforts, the structure of the observed exotic hadrons is still largely unknown and further investigations are needed. In the literature, different interpretations of these hadrons, including tightly-bound  compact tetraquark and pentaquark states, loosely-bound meson-meson and baryon-meson molecular states, charmonium-molecule mixtures or the product of rescattering effects, have been proposed (see, for example, Refs. [13--15]).

Among these exotic hadronic states, the all-heavy tetraquark systems with different flavors have attracted considerable attention as an exotic form of hadronic matter and as ideal systems for exploring tetraquark states [16]. And the first all-heavy tetraquark candidate has been reported in 2020. In this year, the LHCb Collaboration reported the observation of a fully-charmed narrow pronounced tetraquark structure around 6900 MeV, denoted as $X(6900)$
\footnote{$^)$Or as $T_{cc{\bar c}{\bar c}}(6900)^0$ according to the new naming scheme for exotica [17].}$^)$,
in the di-$J/\psi$ invariant mass spectrum, excess of $J/\psi$ pairs just above the ${J/\psi}J/\psi$ threshold
(associated with the narrow near-threshold $cc{\bar c}{\bar c}$ state known as $X(6200)$ [18])
and a small excess around 7200 MeV using proton--proton collision data taken at center-of-mass energies 7, 8 and 13 TeV at the LHC [19]. This structure produced in proton--proton collisions at $\sqrt{s}=13$ TeV was later confirmed by the ATLAS [20] and CMS [21] Collaborations also in the ${J/\psi}J/\psi$ decay channel. More recently, the ATLAS Collaboration has observed as well a clear excess near 6900 MeV in the ${J/\psi}\psi(2S)$ mass spectrum, while the existence of a potential resonance near 7200 MeV, called $X(7200)$, was not supported by the current data [22]. Confirming the $X(6900)$ in the ${J/\psi}\psi(2S)$ mass spectrum, the ATLAS Collaboration measured the ratio $R$ of its partial decay widths between the ${J/\psi}\psi(2S)$ and di-$J/\psi$ channels: $R=1.08\pm0.20^{+0.40}_{-0.17}$ [22]. In the work [21], the CMS Collaboration, in addition to the $X(6900)$, found two new tetracharm states, named $X(6600)$ and $X(7100)$. A clearer picture of these three states was obtained in very recent publication [23] using $pp$ collision data sample collected by the CMS detector and corresponding to an integrated luminosity of 315 fb$^{-1}$, which yields 3.6 times more ${J/\psi}J/\psi$ pairs than previous studies [21] by CMS. The $X(6600/6900/7100) \to {J/\psi}J/\psi$ and $X(6900) \to {J/\psi}\psi(2S)$ decays indicate that they are definitely tetraquark states with minimal quark content $cc{\bar c}{\bar c}$ [23]. Among these states, the $X(6900)$ resonance is of foremost importance as the first distinct fully-charmed tetraquark candidate in the charm sector. Its measured by LHCb [19], assuming the presence of interference between the $X(6900)$ as Breit-Wigner resonance and the non-resonant continuum, mass and width are $m_{X(6900)}=(6886\pm11\pm11$) MeV and $\Gamma_{X(6900)}=(168\pm33\pm69$) MeV, while without this interference they are $m_{X(6900)}=(6905\pm11\pm7$) MeV and $\Gamma_{X(6900)}=(80\pm19\pm33$) MeV. Here, the uncertainties are statistical (first) and systematic (second). The measurements, performed in Refs. [23, 24], strongly favor the $J^{PC}=2^{++}$ quantum numbers assignment for the discovered family of three all-charm tetraquarks $X(6600/6900/7100)$. It should be said that the masses of the $S$-wave fully-charmed $cc{\bar c}{\bar c}$ and fully-bottomed $bb{\bar b}{\bar b}$ tetraquarks have been predicted in the works [25] and [26] before the experimental discovery of $X(6900)$ in 2020 using, respectively, the QCD sum rule within the diquark-antidiquark picture and two non-relativistic quark models. Moreover, the existence of possible fully-heavy tetraquark states has been explored in literature [27--32] even long before this discovery--more than four decades ago. In contrast, experimental evidence for the beauty counterpart $X_{bb{\bar b}{\bar b}}$ of the fully-charmed resonance $X(6900)$ is still lacking. In 2017, the CMS Collaboration reported an indication in the four-muon final state for the existence of the $bb{\bar b}{\bar b}$ exotic meson with a mass around 18.4 GeV studying the $\Upsilon(1S)$ pair production in proton--proton collisions at $\sqrt{s}=8$ TeV [33, 34]. This would set it at 520 MeV below the $\Upsilon(1S){\Upsilon(1S)}$ threshold. Later, the LHCb [35] and CMS [36] Collaborations analyzed the ${\Upsilon(1S)}{\mu^+}\mu^-$ invariant mass distributions from $pp$ collisions at center-of-mass energies 7, 8, 13 TeV and 13 TeV, respectively, with the aim of observation of a possible narrow exotic meson state $X_{bb{\bar b}{\bar b}}$. However, no significant excess of events compatible with this state was observed in data. Theoretically, the $bb{\bar b}{\bar b}$ tetraquark states have been extensively studied in the literature, but the conclusions are controversial. Thus, some works [37--40] support the existence of the stable bound states in the $bb{\bar b}{\bar b}$ system with quantum numbers $J^{PC}=0^{++}, 1^{+-}, 2^{++}$ below the lowest corresponding scattering states, while in others [26, 41--48] no bound states here were found. The numerical results obtained in the latter publications show that the ground $bb{\bar b}{\bar b}$ states are located above the corresponding scattering states.

Experimental observations of the fully-charmed tetraquark states $X(6600/6900/7100)$ have motivated significant efforts
focused on the understanding of their internal structure and production mechanism as well as on the description of the
mass spectra and decay properties of the tetracharm $cc{\bar c}{\bar c}$ system. Thus, the most popular interpretation of the tetraquark state around 6900 MeV of our interest, proposed over the last years, suggests that this state is likely to be the first radial excitation of the $S$-wave $cc{\bar c}{\bar c}$ system with the $J^{PC}=2^{++}$ composed of aligned spin-1  diquark $[cc]$ and antidiquark $[{\bar c}{\bar c}]$ bound by the color force [23, 26, 49--60]
\footnote{$^)$Or the 2$S$ compact diquark-antidiquark tetraquark state according to the spectroscopic notation.
On the contrary, Refs. [54, 61] treat the $X(6900)$ as a $P$-wave $cc{\bar c}{\bar c}$ state. Besides, it was interpreted in Refs. [61, 62] as the second radial excitation of the $cc{\bar c}{\bar c}$ system, or as the 3$S$ tetracharm state.}$^)$. There have been other assignments of the $X(6900)$ state, such as an amorphous system with no internal clustering
which obeys the Pauli principle and where the heavy quarks and antiquarks overlap [63], a gluonic tetracharm hybrid state $cc-G-{\bar c}{\bar c}$ -- a kind of dicharm pair excitation containing a pair of constituent diquarks and a dynamical gluon [64], a dynamically generated resonance pole structure arising in the coupled-channel scattering between vector charmonium pairs [65--69], a structure due to threshold cusps in the lineshapes of the rescatterings into the di-$J/\psi$  all the possible combinations of double charmonia directly produced in a high-energy proton--proton collisions [70, 71], a loosely bound ${J/\psi}J/\psi$ excited molecular cluster [72]. Analysis of the LHCb and CMS data on $X(6900)$, found in the $J/\psi{J/\psi}$ and $J/\psi{\psi(2S)}$ systems, performed in Refs. [66, 73] by using a momentum-dependent Flatte-like parametrization, the pole-counting rule and spectral density function sum rule disfavors the $X(6900)$ structure as a $J/\psi{\psi(2S)}$ molecular state since the $J/\psi{\psi(2S)}$ threshold is far away (by approximately
100 MeV) from $X(6900)$ mass and points out that there still exists the possibility that the $X(6900)$ is a molecular state composed of virtual charmonium pairs whose mass thresholds are closer to it, such as $J/\psi{\psi(3770)}$, $J/\psi{\psi_2(3823)}$, $J/\psi{\psi_3(3842)}$, ${\chi_{c0}}{\chi_{c2}}$ pairs bound by the residual strong interaction [58, 59, 74]
\footnote{$^)$In addition, Ref. [75] indicates that the attraction between $J/\psi$ and $\psi(3770)$ has to be strong enough to form the molecular resonant structure $X(6900)$ and argues that since the $\psi(3770)$ decays mainly into
the $D{\bar D}$ pair, the interactions between $J/\psi$ and $\psi(3770)$ may be significantly enhanced due to the three point $D{\bar D}$ loop diagram.}$^)$.
Similarly, the resonance $X(6900)$ may be interpreted as hadronic molecules ${\bar \chi}_{c0}{\chi_{c0}}$ and
${\chi_{c0}}{\chi_{c0}}$, as was argued in Refs. [76] and [77], respectively
\footnote{$^)$It is also worth noting that another tetracharm near-threshold exotic state $X(6200)$ was modeled
in Refs. [68] and [77, 78], correspondingly, as a $2^{++}$ di-$J/\psi$ and as an $0^{++}$ ${\eta_c}\eta_c$ molecules. In this context, Ref. [79] pointed out that the di-$J/\psi$ femtoscopic correlation function may serve as a clear experimental observable to distinguish the nature of $X(6200)$ -- whether it is a resonant, bound, or virtual state. }$^)$.
Based on the above, it is clear that, despite a lot of theoretical and experimental efforts, the deep understanding of the nature of exotic $X(6900)$ state (and other observed fully-charmed tetraquark states) is still lacking and more further efforts, especially from the experimental side,
are needed.

To get further insight into the nature of exotic $X(6900)$ state, it is of importance to study its photoproduction on nuclei at energies close to the threshold for its production off a free nucleon. This has the advantage compared to the high-energy hadronic collisions  that the interpretation of data from photoproduction experiments on nuclei is clearer in particular due to a negligible strength of initial-state photon interaction.
The study of the photoproduction of $X(6900)$ state off a proton target at near-threshold energies has been carried out previously in Ref. [80]. Its exclusive total production cross section from $\omega$ meson exchange has been predicted here. Fully-charmed tetraquark production in $pp$, $pA$ and $AA$ collisions at LHC energies in the kinematic ranges accessible at the ongoing LHC experiments has been theoretically studied in Refs. [81--87], [86, 87] and [87], respectively. Also, fully-charmed tetraquark production in electron--proton and electron--positron collisions has been investigated in Refs. [88, 89] and [90], correspondingly. In this work we present the detailed predictions of absolute and relative observables for the $X(6900)$ photoproduction off $^{12}$C and $^{184}$W target nuclei at threshold energies obtained in the framework of the collision model, based on the nuclear spectral function, within three different scenarios for its internal structure. The predictions can be confronted to the experimental data from the future measurements at the electron-ion colliders EIC [91] and EicC [92] to discriminate between these scenarios.

\section*{2. Formalism: Direct $X(6900)$ photoproduction mechanism}

\hspace{1.5cm} Direct photoproduction of a narrow exotic resonant state $X(6900)$ with mass $m_{X(6900)}=6.886$ GeV and with quantum numbers $J^{PC}=2^{++}$ [19, 23, 24] on nuclear targets at the near-threshold laboratory incident photon energies $E_{\gamma} \le 40$ GeV of interest
\footnote{$^)$Which corresponds to the center-of-mass energies $W$ of the photon-proton system $W \le 8.72$ GeV,
or to the relatively "low" excess energies $\epsilon$ above the $X(6900)p$ production threshold
$0 \le \epsilon \le 0.9$ GeV and where the $X(6900)$ mesons can be observed in the ${\gamma}p$ and ${\gamma}A$ reactions at the under-construction electron-ion colliders EIC [91] and EicC [92] in the United States and China.}$^)$
may proceed via the following elementary processes with the lowest free production threshold ($\approx$ 32.154 GeV)
[80]:
\begin{equation}
{\gamma}+p \to X(6900)+p,
\end{equation}
\begin{equation}
{\gamma}+n \to X(6900)+n.
\end{equation}
The $X(6900)$ mesons and nucleons, produced in these processes, are sufficiently energetic (see, Fig. 1).
Thus, for example, the kinematically allowed $X(6900)$ meson with the nominal mass $m_{X(6900)}=6.886$ GeV
measured by LHCb [19] and final proton laboratory momenta in the direct process (1), proceeding on the free target proton at rest, vary within the momentum ranges of 26.251--33.219 GeV/c and 1.781--8.749 GeV/c, respectively, at incident photon energy of $E_{\gamma}=35$ GeV.
\begin{figure}[htb]
\begin{center}
\includegraphics[width=15.0cm]{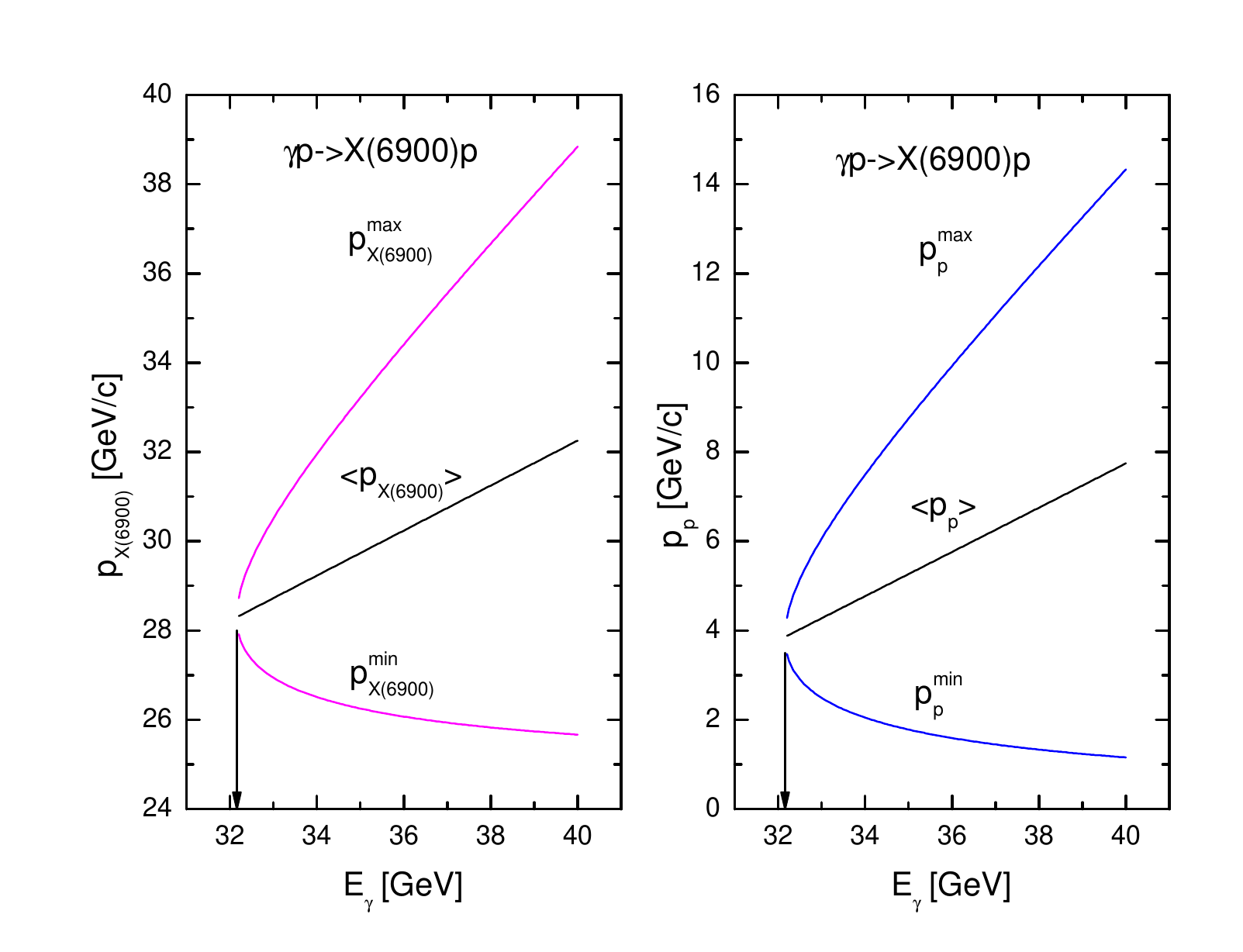}
\vspace*{-2mm} \caption{(Color online.) The kinematically allowed $X(6900)$ meson with the nominal mass $m_{X(6900)}=6.886$ GeV and final proton maximal and minimal laboratory momenta, their arithmetic averages, $<p_{X(6900)}>$ and $<p_p>$, in the direct ${\gamma}p \to X(6900)p$ process, taking place on the free target proton being at rest, as functions of the laboratory energy $E_{\gamma}$ of the photon--proton collisions in the considered kinematic range (respectively, left and right panels). The arrows indicate the threshold energy of 32.154 GeV for the $X(6900)$
photoproduction with the nominal mass $m_{X(6900)}=6.886$ GeV on a free nucleon.}
\label{void}
\end{center}
\end{figure}
\begin{figure}[htb]
\begin{center}
\includegraphics[width=15.0cm]{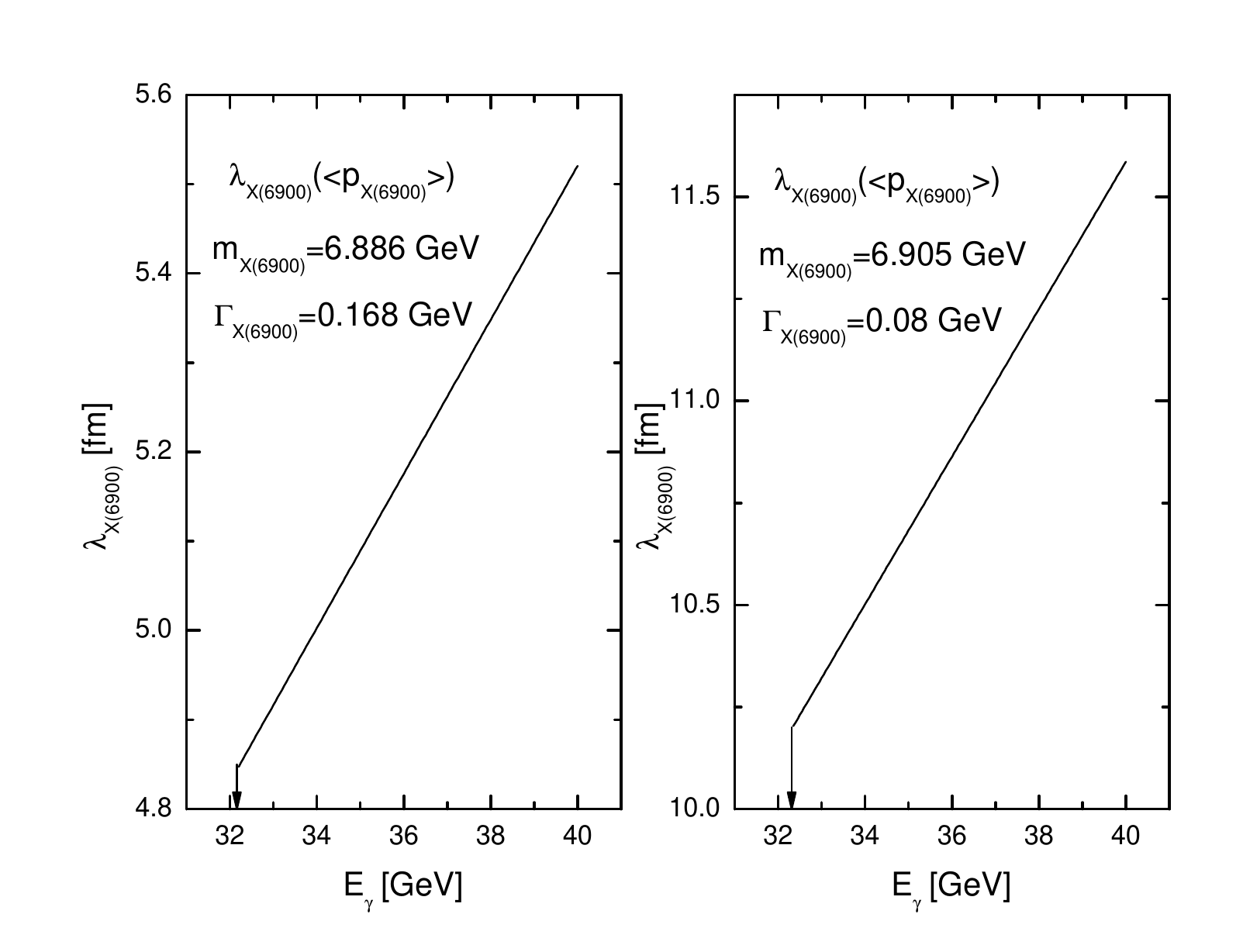}
\vspace*{-2mm} \caption{(Color online.) The decay mean free path of the $X(6900)$ resonance with the laboratory momentum $<p_{X(6900)}>$, presented in the left panel of Fig. 1, and with the nominal masses and total decay widths in its rest frame measured by LHCb [19]: $m_{X(6900)}=6.886$ GeV, $\Gamma_{X(6900)}=0.168$ GeV (left panel); $m_{X(6900)}=6.905$ GeV, $\Gamma_{X(6900)}=0.08$ GeV (right panel) as a function of the laboratory energy $E_{\gamma}$ of the photon--proton collisions in the considered kinematic range. The arrows indicate the threshold energies of 32.154 GeV and 32.313 GeV for the $X(6900)$ photoproduction with these masses on a free nucleon.}
\label{void}
\end{center}
\end{figure}
The kinematical characteristics of particles, produced in the reaction (2), and produced again in reactions (1), (2) when the nominal mass of $X(6900)$ is taken to be $m_{X(6900)}=6.905$ GeV [19], as our calculations have shown, are quite similar to those of final particles in the process (1) depicted in Fig. 1. Since the medium effects are expected to be reduced for high momenta, we will ignore the medium modifications of the outgoing sufficiently energetic $X(6900)$ mesons and nucleons in the case when the reactions (1), (2) proceed on a nucleons embedded in a nuclear target
\footnote{$^)$It should be pointed out that the behavior of fully-heavy charmed and bottomed tetraquarks in hot strongly interacting medium has been studied in the recent works [93, 94]. In them, the temperature dependence of their masses
and dissociation temperatures was calculated. In particular, in Ref. [94] it was obtained using QCD sum rule calculations based on a diquark-antidiquark picture for the ground state of fully-heavy resonances that around the critical temperature the mass of the $cc{\bar c}{\bar c}$ state decreases by 8\% compared to its vacuum value, while for its bottom partner, this percentage is approximately 3.3\%.}$^)$.

Then, neglecting the distortion  of the incident photon in the nucleus and the attenuation of the $X(6900)$
flux here due to its decay
\footnote{$^)$Most of the $X(6900)$'s decay essentially outside the target nuclei of interest. This is due to
the following. The $X(6900)$ resonance decay mean free path can be estimated as
$\lambda_{X}=p_{X}/(m_{X}\Gamma_{X})$, where $X$ stands for $X(6900)$, and $p_{X}$ and $\Gamma_{X}$ are its
laboratory momentum and vacuum total decay width in its rest frame. For average values $p_{X}=<p_{X(6900)}>$,
presented in the left panel of Fig. 1, and for the widths (and masses, see above) $\Gamma_{X}=0.168$ GeV and $\Gamma_{X}=0.08$ GeV, based on the interference and no-interference with the background fits [19], we have in line with Fig. 2 that $\lambda_{X}$ $\sim$ 5 fm and 11 fm, respectively, in the photon energy range of interest. The first value is larger than the radius of $^{12}$C, which is approximately 3 fm, and is comparable to that of $^{184}$W of about 7.4 fm, whereas the second value is larger than the radii of both these target nuclei. Due to this fact, to reduce the uncertainty of the theoretical predictions for the $X(6900)$ yield from nuclei it is natural, in view of large uncertainties in the measured in [19] $X(6900)$ width and in that extracted in [95] from a combined analysis of the LHCb-ATLAS-CMS di-$J/\psi$ mass spectrum using four different fit models, to neglect the $X(6900)$ decays inside them.}$^)$
as well as describing the $X(6900)$ absorption by target nucleons at the expense of the inelastic $X(6900)$--nucleon collisions by the absorption cross section $\sigma_{X(6900)N}$, we represent the total cross section for the production
of $X(6900)$ mesons with quantum numbers $J^{PC}=2^{++}$ on nuclei from the direct photon--induced reaction
channels (1), (2) as follows [96]:
\begin{equation}
\sigma_{{\gamma}A\to X(6900)X}^{({\rm dir})}(E_{\gamma})=I_{V}[A,\sigma_{X(6900)N}]
\left<\sigma_{{\gamma}p \to X(6900)p}(E_{\gamma})\right>_A,
\end{equation}
where the effective number of target nucleons participating in the direct processes (1), (2),\\ $I_{V}[A,\sigma_{X(6900)N}]$, and "in-medium" total cross section for the production of $X(6900)$ mesons with
quantum numbers $J^{PC}=2^{++}$ in reaction (1) $\sigma_{{\gamma}p \to X(6900)p}(\sqrt{s^*})$ at the in-medium ${\gamma}p$ center-of-mass energy $\sqrt{s^*}$, averaged over target nucleon binding and Fermi motion, $\left<\sigma_{{\gamma}p \to X(6900)p}(E_{\gamma})\right>_A$, are defined by Eqs. (4), (5) and (6) from Ref. [96], respectively, in which one needs to make the substitution: $X(3872) \to X(6900)$
\footnote{$^)$In Eq. (3) we assume that the $X(6900)$ meson production cross sections in reactions (1) and (2)
are the same and neglect the difference between proton ($m_p$) and neutron ($m_n$) masses. For target nuclei of interest, the nuclear spectral function, used in our calculations of "in-medium" total cross section
$\left<\sigma_{{\gamma}p \to X(6900)p}(E_{\gamma})\right>_A$ entering into this equation, was taken from Ref. [97].}$^)$.

  As before in Ref. [96], we suggest here that the "in-medium" cross section
$\sigma_{{\gamma}p \to X(6900)p}({\sqrt{s^*}})$ for $X(6900)$ production with quantum numbers $J^{PC}=2^{++}$
in reaction (1) is equivalent to the vacuum cross section $\sigma_{{\gamma}p \to X(6900)p}({\sqrt{s}})$,
in which the free space center-of-mass energy squared $s$ for given photon laboratory energy $E_{\gamma}$
and momentum ${\bf p}_{\gamma}$, presented by the formula
\begin{equation}
s=s(E_{\gamma})=W^2=(E_{\gamma}+m_p)^2-{\bf p}_{\gamma}^2=m_p^2+2m_pE_{\gamma},
\end{equation}
is replaced by the in-medium expression
\begin{equation}
  s^*=(E_{\gamma}+E_t)^2-({\bf p}_{\gamma}+{\bf p}_t)^2,
\end{equation}
\begin{equation}
   E_t=M_A-\sqrt{(-{\bf p}_t)^2+(M_{A}-m_{p}+E)^{2}}.
\end{equation}
Here, $E_t$, ${\bf p}_{t}$ and $E$ are the total energy, momentum and binding energy of the struck target
proton involved in the collision process (1).
Presently, in the literature there are only some theoretical predictions
for the free total cross section for the $X(6900)$ production with quantum numbers $J^{PC}=0^{++}$ in the reaction (1)
at the considered photon energies $E_{\gamma} \le $ 40 GeV (cf. Ref. [80]). At these energies,
for this cross section, denoted by the $\sigma_{{\gamma}p \to {\tilde X}(6900)p}(\sqrt{s})$,
we have used the following parametrization based on its behavior for energies near threshold
\begin{equation}
\sigma_{{\gamma}p \to {\tilde X}(6900)p}(\sqrt{s})=a\left(1-\frac{s_{\rm th}}{s}\right)^{b},
\end{equation}
where for $m_{X(6900)}=6.886$ GeV
\begin{equation}
s_{\rm th}=(m_{X(6900)}+m_{p})^2=(7.82428~{\rm GeV})^2
\end{equation}
and free parameters $a$ and $b$ were obtained by fitting with (7) the results of its calculations in Ref. [80], shown
here in Fig. 6, within the $\omega$ meson exchange model assuming a nominal branching fraction $Br[{\tilde X}(6900) \to {J/\psi}\omega]=1$~\% at the center-of-mass energies of 8 and 9 GeV of the photon-proton system. The resulting values
of these parameters are $a=2.253$ nb and $b=0.576$.
In passing forward, we note that in the energy regime $E_{\gamma} \le 40$ GeV, the cross section (7) can reach  the values of the order of 1 nb. The cross section $\sigma_{{\gamma}p \to {X(6900)}p}(\sqrt{s})$ for free space production of the $X(6900)$ with quantum numbers $J^{PC}=2^{++}$ in the process (1) was evaluated by us assuming the consistency of the cross section ratio between the $2^{++}$ and the $0^{++}$ states, $\sigma_{{\gamma}p \to {X(6900)}p}(\sqrt{s})/\sigma_{{\gamma}p \to {\tilde X}(6900)p}(\sqrt{s})$, with the simple spin-state counting rule of 5/1 [81--83, 98]
\footnote{$^)$It is worth mentioning that the approach [86] predicts the following estimate of the production cross section $\sigma_{T_{4c}}$ of fully-charmed tetraquark, dubbed $T_{4c}$, in high-energy hadronic collisions knowing the
one $\sigma_{X(3872)}$ for X(3872) in these collisions: $\sigma_{T_{4c}}{\approx}~0.12\sigma_{X(3872)}$.
For the time being, we assume that this estimate is also correct for ${\gamma}p$ collisions at threshold energies.
However, in it the former and latter cross sections are calculated, respectively, at the c.m. energies $W$ and
${\tilde W}$, which correspond to the same excess energies above the $X(6900)p$ and $X(3872)p$ thresholds. Using, for example, the value of $\sigma_{X(3872)}$${\approx}$~30~nb predicted in [80] for the collision energy of ${\tilde W}=5.8$
GeV, corresponding to the excess energy above the $X(3872)p$ threshold of 1 GeV (see Fig. 3 in [80]), we have that
$\sigma_{T_{4c}}$${\approx}$3.6~nb. On the other hand, the total cross section of the ${\gamma}p \to X(6900)p$ reaction
for $X(6900)$ with the spin-parity assignment $J^{PC}=0^{++}$ is of about 0.8 nb at the c.m. energy $W=8.8$ GeV,
also corresponding to the excess energy above the $X(6900)p$ threshold of 1 GeV (see Fig. 6 in [80]).
In view of Refs. [81-83, 98] (and Ref. [119]), we assume in our work that the $J=2$ tetraquarks are produced
in ${\gamma}p$ collisions with a statistical factor of 5 with respect to the spin 0 state. With these, we have that the
cross section for producing $X(6900)$ with the $J^{PC}=2^{++}$ amounts approximately 4 nb. This value is in a good
agreement with that of 3.6 nb reported above for the $T_{4c}$ tetraquark. This fact may give a confidence to us that the
predicted in [80] the $J^{PC}=0^{++}~X(6900)$ near-threshold photoproduction cross sections on a proton target
within the $\omega$ meson exchange model assuming a nominal branching fraction $Br[X(6900)~{\rm with}~J^{PC}=0^{++} \to {J/\psi}\omega]=1$~\% and the adopted assumption that the ratio $\sigma(2^{++})/\sigma(0^{++})=5$ are pretty realistic.
This means that we can definitely use the parametrization (7) and the cross section ratio $\sigma(2^{++})/\sigma(0^{++})=5$ to provide certain guidance for the measurements of the absolute yields of X(6900)
mesons from nuclei in future photoproduction experiments and to impose some (preliminary) constraints on their internal
structure. To put a strong constraints on this structure using the $X(6900)$ absolute yields one needs to know precisely
the proton (neutron)-target $X(6900)$ production cross sections.}$^)$.

The local proton and neutron densities, adopted in the calculations of the quantity $I_{V}[A,\sigma_{{X(6900)}N}]$, entering into Eq. (3), for the target nuclei $^{12}_{6}$C, $^{27}_{13}$Al, $^{40}_{20}$Ca, $^{63}_{29}$Cu, $^{93}_{41}$Nb, $^{112}_{50}$Sn, $^{184}_{74}$W, $^{208}_{82}$Pb and $^{238}_{92}$U considered in the present work
are given in Ref. [96]. For medium-weight $^{93}_{41}$Nb, $^{112}_{50}$Sn and heavy $^{184}_{74}$W, $^{208}_{82}$Pb, $^{238}_{92}$U target nuclei we use the neutron density $\rho_n(r)$ in the 'skin' form.

To estimate the rate of the $2^{++}$ $X(6900)$ meson photoproduction on nuclei, one needs also to specify its absorption by nucleons cross section $\sigma_{X(6900)N}$, which determines the quantity $I_{V}[A,\sigma_{X(6900)N}]$ (cf. Eqs. (4), (5) from Ref. [96]). With this aim, we consider in our study two different popular scenarios for the $X(6900)$ discussed above: i) compact, $\sim$ 1 fm, diquark-antidiquark $[cc]$-$[{\bar c}{\bar c}]$ 2$S$-wave tetraquark state with the $J^{PC}=2^{++}$, ii) ${J/\psi}\psi(3770)$ hadronic molecule
\footnote{$^)$Since the ${J/\psi}\psi(3770)$ threshold (6870.6 MeV) is closer to the measured by the LHCb (in the interference scenario) [19], ATLAS (in the fit Model A) [20] and CMS (in the ${J/\psi}\psi(2S)$ mode: Run 2+3) [23] Collaborations central masses of the $X(6900)$ state (respectively, 6886 MeV, 6860 MeV and 6876 MeV) than those of
the $J/\psi{\psi(2S)}$, $J/\psi{\psi_2(3823)}$, $J/\psi{\psi_3(3842)}$ and ${\chi_{c0}}{\chi_{c2}}$ charmonium pairs
(correspondingly, 6783 MeV, 6919.1 MeV, 6939.6 MeV and 6970.88 MeV) [17], it is natural to assume that the ${J/\psi}\psi(3770)$ "molecule", in which two mesons are stuck together loosely and are weakly interacting through the residual strong force, may be a good candidate for the $X(6900)$ molecular state.}$^)$.
In addition, to extend the range of applicability of our model, we will also adopt in our calculations another additional option for the $X(6900)$, namely: hybrid state -- state in which the $X(6900)$ is considered as a mixture of the compact diquark-antidiquark and molecular components. And this mixture contains 50\% of the genuine non-molecular 4$c$ component and 50\% of the molecular ${J/\psi}\psi(3770)$ component.

Interpreting the $X(6900)$ as a compact tetraquark state with radius $r_{4c}$, it is natural to approximate the high-momentum  $X(6900)$--nucleon absorption cross section $\sigma_{{X(6900)}N}^{{\rm 4{c}}}$ in this picture
by a geometrical cross section $\sigma^{\rm geo}_{4c}=\pi{r_{4c}^2}$, {\it i.e.}, $\sigma_{{X(6900)}N}^{{\rm 4{c}}}\approx{\pi}r_{4c}^2$. Geometrical considerations show that the radius $r_{4c}$ can be represented as one half
of the sum of the r.m.s. size $r_{cc}$ of the interquark distance in diquark and the r.m.s. size $r_{[cc][{\bar c}{\bar c}]}$ of the distance between the diquark and antidiquark inside the $2^{++}$ 2$S$-wave tetraquark.
For $r_{cc}=0.55$ fm and $r_{[cc][{\bar c}{\bar c}]}=0.60$ fm [54], we obtain that $r_{4c}=0.575$ fm (cf. Ref. [99]) and $\sigma_{{X(6900)}N}^{{\rm 4{c}}}=10.4$ mb. On the other hand, the radius $r_{4c}$ can be expressed as one half
of the sum of the r.m.s. size $r_{c{\bar c}}$ of the distance between quark and antiquark and the r.m.s. size $r_{[c{\bar c}][c{\bar c}]}$ of the distance between the quark-antiquark pairs inside the $2^{++}$ 2$S$-wave tetraquark.
For $r_{c{\bar c}}=0.72$ fm and $r_{[c{\bar c}][c{\bar c}]}=0.39$ fm [54], we get that $r_{4c}=0.555$ fm and $\sigma_{{X(6900)}N}^{{\rm 4{c}}}=9.7$ mb. With these, for the cross section $\sigma_{{X(6900)}N}^{{\rm 4{c}}}$ we take
an arithmetic average of the obtained results: $\sigma_{{X(6900)}N}^{{\rm 4{c}}}=10$ mb.

Within the adopted "pure" hadronic molecular interpretation of the $X(6900)$ as the $J/\psi{\psi(3770)}$ weak bound state, we assume that this state dissociates while one of its constituent charmonia taken to be on-shell and flying with an average laboratory momentum $\sim$ 15 GeV/c (see above) quasi-free scatters (elastically or inelastically) with the intranuclear nucleon (proton or neutron) and the other charmonium is a spectator (cf. Ref. [96])
\footnote{$^)$It should be noted that this assumption is sufficiently well justified for meson--meson molecular states
with typical inter-meson distances $\ge 1$ fm [99]. In this case the constituents can be considered as on-shell individual particles flying together and interactions between them can be ignored. Thus, using the expression $r_{X(6900)}=1/\sqrt{4\mu_0|\delta_{X(6900)}|}$ for the r.m.s. size $r_{X(6900)}$ of the inter-meson distance [96], where $\mu_0$ is the $J/\psi{\psi(3770)}$ reduced mass and $\delta_{X(6900)}$ is "the binding energy" $\delta_{X(6900)}=m_{J/\psi}+m_{\psi(3770)}-m_{X(6900)}=6870.6$ MeV-$m_{X(6900)}$ of the molecule,
and taking, for example, $m_{X(6900)}=6876$ MeV [23], we find that $\delta_{X(6900)}=-5.4$ MeV and the charmonium mesons in the $X(6900)$ have a r.m.s. separation: $r_{X(6900)}=1.03$ fm. This size scale enables to consider the $J/\psi$ and
$\psi(3770)$ mesons having a mean radii $\sim$ 0.2 fm and 0.45 fm, respectively, (see below) as not overlapping individual particles retaining their own identities. This implies that it may be sufficiently correctly to treat the $X(6900)$ interaction with the target nucleons in the manner adopted in the present work (see Eq. (9)). }$^)$.
Then, in this quasi-free approximation we can evaluate the cross sections for $X(6900)$ absorption in the molecular scenario, $\sigma_{X(6900)N}^{\rm mol}$, as:
\begin{equation}
\sigma_{{X(6900)}N}^{\rm mol}\approx\sigma_{{J/\psi}N}+\sigma_{\psi(3770)N},
\end{equation}
Here, $\sigma_{{J/\psi}N}$ and $\sigma_{{\psi(3770)}N}$ are the ${J/\psi}N$ and ${\psi(3770)}N$ absorption cross
sections, respectively. For the $J/\psi$--nucleon absorption cross section $\sigma_{{J/\psi}N}$ we have used the
value $\sigma_{{J/\psi}N}=3.5$ mb motivated by the results from the $J/\psi$ photoproduction experiment at SLAC [100].
Information on the $\psi(3770)$--nucleon dissociation cross section is scarce in the literature. Thus, the quark interchange between a nucleon and a $c{\bar c}$ meson model [101] gives the value of the order of 10$^{-2}$ mb
for this cross section for the laboratory $\psi(3770)$ momenta relevant for the present study. For our opinion, this value is unrealistic enough since the $\psi(3770)$--nucleon dissociation cross section is expected to be greater than that of the ${J/\psi}N$ due to the larger spatial size -- the r.m.s. $c{\bar c}$ separation -- of the $\psi(3770)$ ($\sim$ 0.9 fm [102, 103]) compared to that of the $J/\psi$ ($\sim$ 0.4 fm [99, 104]). Theoretically,
the assumption that the charmonium--nucleon total cross section scales with the charmonium spatial size squared
[105--107] leads to the conclusion that $\psi(3770)N$ total cross section, in view of the above, is  a factor
of about 5.06 larger than that of the ${J/\psi}N$. This leads to its value of about 17.7 mb for the $J/\psi$--nucleon cross section $\sigma_{{J/\psi}N}=$ 3.5 mb [100]. Using the above values for the cross sections $\sigma_{{J/\psi}N}$ and $\sigma_{{\psi(3770)}N}$, we obtain that $\sigma_{{X(6900)}N}^{\rm mol}=21.2$ mb.

In the hybrid scenario, it is assumed that the $X(6900)$ wave function is a linear superposition of the compact four-quark (or 4$c$) and molecular components:
\begin{equation}
|X(6900)>_{\rm hyb}=\alpha|[cc][{\bar c}{\bar c}]>+{\beta}|{J/\psi}\psi(3770)>,
\end{equation}
where, in line with the preceding, $\alpha^2=\beta^2=0.5$. In the hybrid interpretation (10) of the $X(6900)$, we can represent  the $X(6900)$--nucleon absorption cross sections $\sigma_{{X(6900)}N}^{\rm hyb}$ in the following incoherent probability-weighted sum [108]:
\begin{equation}
\sigma_{X(6900)N}^{\rm hyb}=\alpha^2\sigma_{X(6900)N}^{\rm 4{c}}+\beta^2\sigma_{X(6900)N}^{\rm mol}.
\end{equation}
According to the above, we set $\sigma_{X(6900)N}^{\rm 4{c}}=10$ mb and $\sigma_{X(6900)N}^{\rm mol}=21.2$
mb. With these values, the $\sigma_{X(6900)N}^{\rm hyb}$ absorption cross section (11) is $\sigma_{X(6900)N}^{\rm hyb}=15.6$ mb for the non-molecular and molecular probabilities of the $X(6900)$ 50\% and 50\%.
In the end, we summarize the results obtained above for the $X(6900)$--nucleon absorption cross section $\sigma_{X(6900)N}$ in the adopted scenarios for its intrinsic configuration:
\begin{equation}
\sigma_{X(6900)N}=\left\{
\begin{array}{lll}
	10.0~{\rm mb}
	&\mbox{for compact 4c state}, \\
	&\\
    15.6~{\rm mb}
	&\mbox{for hybrid state (50\%,50\%)},\\
	&\\
    21.2~{\rm mb}
	&\mbox{for ${J/\psi}\psi(3770)$ molecule}.
\end{array}
\right.	
\end{equation}
We can see that the $X(6900)$ as the ${J/\psi}\psi(3770)$ molecule has the largest absorption cross section and,
therefore, it is expected to be more easily absorbed in this case than its other internal configurations in a nuclear medium. We will employ these values for the quantity $\sigma_{X(6900)N}$ in our subsequent calculations.

As a measure for the $X(6900)$ absorption cross section $\sigma_{X(6900)N}$ in nuclei (and, hence, for its intrinsic
structure) we will use two additional integral observables.  The first one is the so-called $X(6900)$ transparency ratio defined as [109--112]:
\begin{equation}
S_A=\frac{\sigma_{{\gamma}A \to X(6900)X}^{({\rm dir})}(E_{\gamma})}{A~\sigma_{{\gamma}p \to X(6900)p}(\sqrt{s(E_{\gamma})})},
\end{equation}
{\it i.e.} the ratio between the inclusive nuclear $X(6900)$ photoproduction cross section (3) and by $A$ times the same quantity on a free proton. The second one is the $X(6900)$ transparency ratio $S_A$ normalized to a light nucleus
like $^{12}$C [110, 111, 113]:
\begin{equation}
T_A=\frac{S_A}{S_{\rm C}}=\frac{12~\sigma_{{\gamma}A \to X(6900)X}^{({\rm dir})}(E_{\gamma})}{A~\sigma_{{\gamma}{\rm C} \to X(6900)X}^{({\rm dir})}(E_{\gamma})}.
\end{equation}
The quantities (13) and (14) are the cross-section ratios. They  are sensitive to the $X(6900)$--nucleon absorption cross section. On the other  hand, they are less sensitive than cross sections themselves to the theoretical uncertainties associated mainly with the experimentally unknown total cross sections of the elementary reactions (1), (2).
\begin{figure}[!h]
\begin{center}
\includegraphics[width=15.0cm]{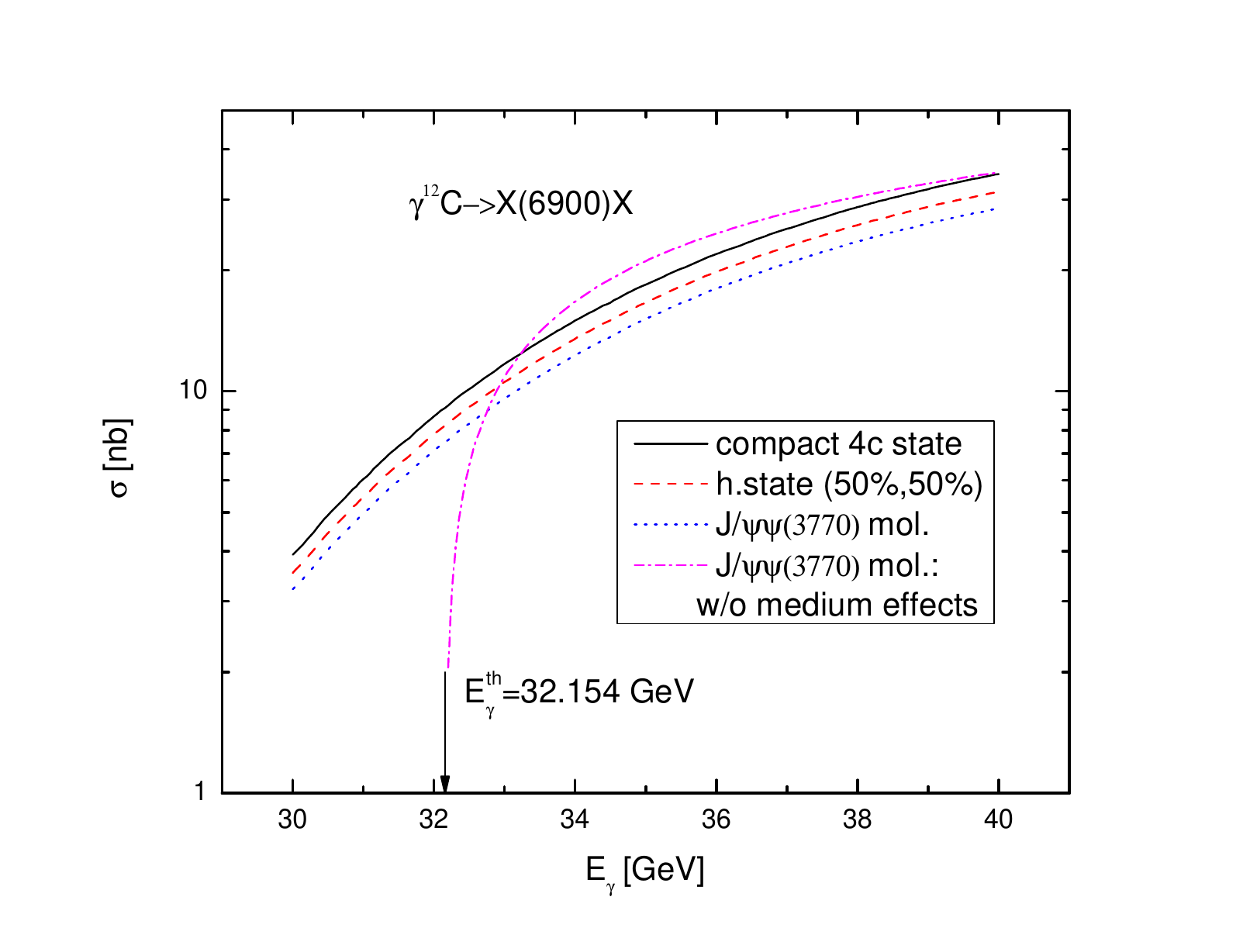}
\vspace*{-2mm} \caption{(Color online.) Excitation function for production of $X(6900)$
mesons with quantum numbers $J^{PC}=2^{++}$ off $^{12}$C from the direct reactions (1), (2) proceeding on an off-shell target nucleons and on a free ones being at rest. The curves are calculations in the scenarios, in which the $X(6900)$ is treated as a purely compact four-quark (4$c$) state, as a purely ${J/\psi}\psi(3770)$ molecular state, or as a hybrid state: mixture of the non-molecular (compact) and molecular (non-compact) components, in which there is 50\% of the 4$c$ component and 50\% molecular component. The arrow indicates the threshold energy for the $X(6900)$
photoproduction on a free nucleon.}
\label{void}
\end{center}
\end{figure}
\begin{figure}[!h]
\begin{center}
\includegraphics[width=15.0cm]{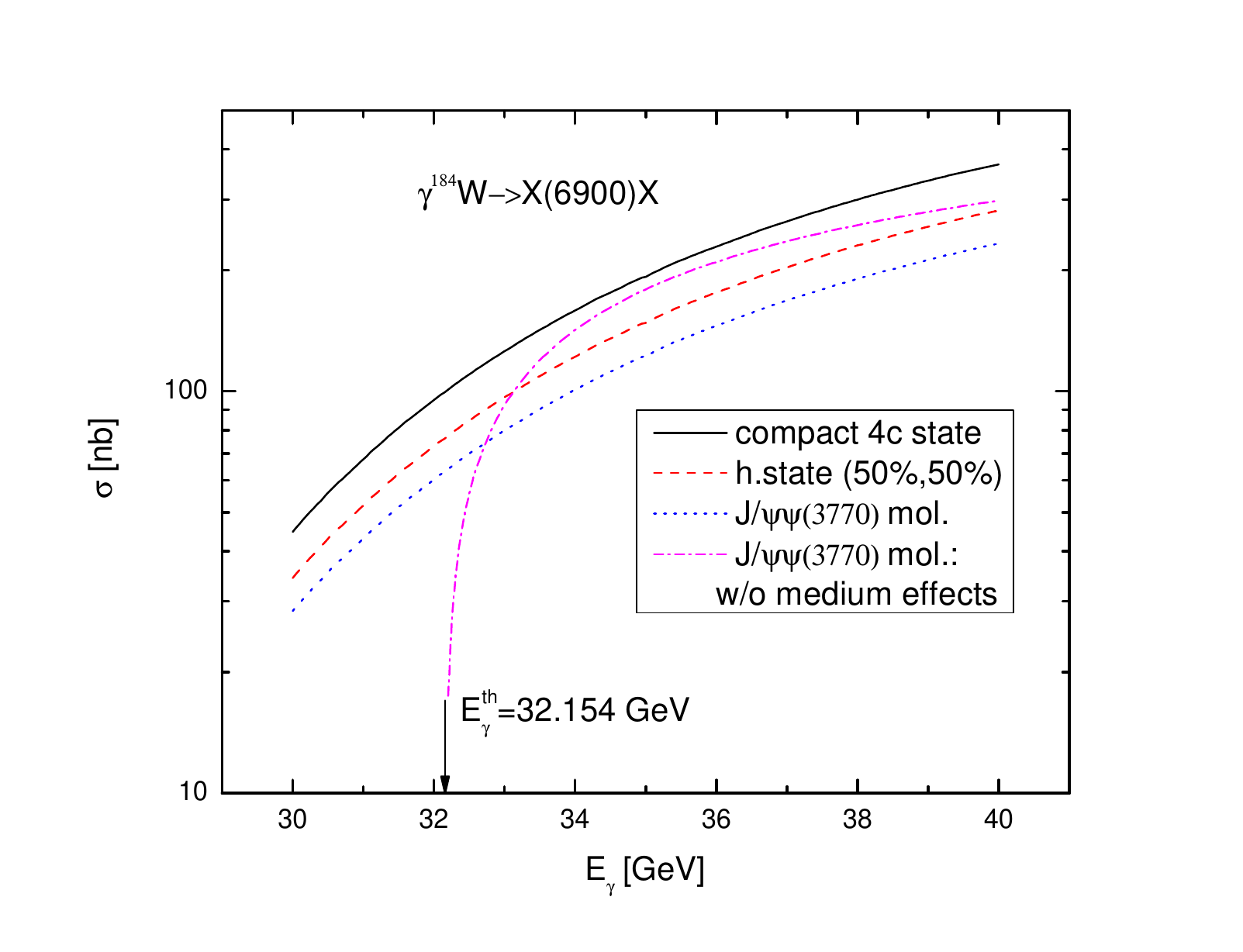}
\vspace*{-2mm} \caption{(Color online.) The same as in Fig. 3, but for the $^{184}$W target nucleus.}
\label{void}
\end{center}
\end{figure}

The information on the $X(6900)$ absorption cross section $\sigma_{X(6900)N}$ (and on its internal structure)
can also be extracted from the comparison of the measured and calculated momentum distributions of $X(6900)$ mesons from nuclei in the photon energy range of interest.
Therefore, we consider now the momentum-dependent inclusive differential cross section for their production
with momentum $p_{X(6900)}$ from the direct processes (1) and (2) in ${\gamma}A$ reactions.
Due to the fact that the $X(6900)$ meson moves in the nucleus essentially forward in the lab frame
\footnote{$^)$Thus, for example, the maximum angle of its production on a free target proton at rest in
reaction (1) is about 1.5$^{\circ}$ at photon energy of 35 GeV.}$^)$,
we will calculate the $X(6900)$ momentum distribution from the considered target nuclei
for the laboratory solid angle ${\Delta}{\bf \Omega}_{X(6900)}$ = $0^{\circ} \le \theta_{X(6900)} \le 5^{\circ}$,
and $0 \le \varphi_{X(6900)} \le 2{\pi}$. Then, according to the results presented both in Ref. [96] and above by
Eq. (3), we can get the following expression for this distribution:
\begin{equation}
\frac{d\sigma_{{\gamma}A\to {X(6900)}X}^{({\rm dir})}
(p_{\gamma},p_{X(6900)})}{dp_{X(6900)}}=
2{\pi}I_{V}[A,\sigma_{X(6900)N}]
\int\limits_{\cos5^{\circ}}^{1}d\cos{{\theta_{X(6900)}}}
\left<\frac{d\sigma_{{\gamma}p\to {X(6900)}{p}}(p_{\gamma},
p_{X(6900)},\theta_{X(6900)})}{dp_{X(6900)}d{\bf \Omega}_{X(6900)}}\right>_A,
\end{equation}
where
$\left<\frac{d\sigma_{{\gamma}p \to {X(6900)}p}(p_{\gamma},
p_{X(6900)},\theta_{X(6900)})}{dp_{X(6900)}d{\bf \Omega}_{X(6900)}}\right>_A$
is the off-shell differential cross section for production of $X(6900)$ mesons
with momentum ${\bf p}_{X(6900)}$ in the process (1),
averaged over the Fermi motion and binding energy of the intranuclear protons.
It can be expressed by Eqs. (28), (31)--(39) from Ref. [114], in which one needs to make the
substitution: $\Upsilon(1S) \to X(6900)$. For the sake of a clear presentation of this work, we do not give
these expressions here. In order to calculate the c.m. $X(6900)$ angular distribution in process (1)
(cf. Eq. (34) from Ref. [114]) one needs to know its exponential $t$-slope parameter $b_{X(6900)}$ in
the threshold energy region. Unfortunately, nothing is known so far about it. As a temporary work-around, for the
time being, accounting for the similarity between the fully-charmed tetraquarks and the ordinary charmonia,
we adopt for this parameter the value of 3.0 GeV$^{-2}$ corresponding [114] to the $J/\psi$ slope parameter $b_{J/\psi}$ in the reaction ${\gamma}p \to {J/\psi}p$ at incident photon energy of 35 GeV. Such choice of the $X(6900)$ slope parameter $b_{X(6900)}$ enables us  to estimate the absolute momentum distributions of the produced tetraquark and their ratios (see, respectively, Figs. 5 and 7 given below). Evidently, an analysis of these ratios with the aim of obtaining a definite information about the inner structure of $X(6900)$ has the advantage compared to the analysis of the absolute differential cross sections that they weakly depend both on the experimentally unknown total cross sections of the elementary reactions (1), (2) and on the choice of the slope parameter $b_{X(6900)}$.
\begin{figure}[!h]
\begin{center}
\includegraphics[width=15.0cm]{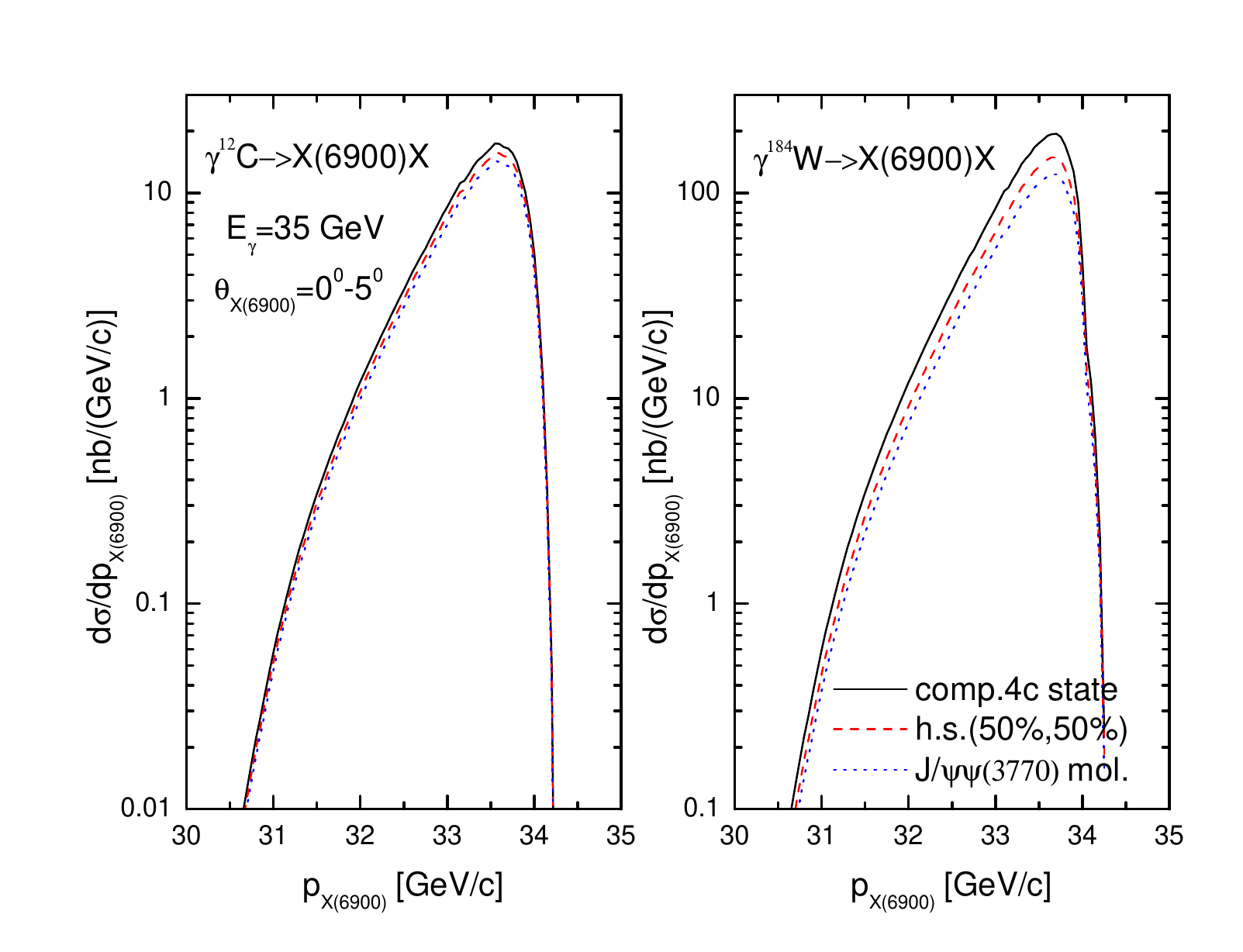}
\vspace*{-2mm} \caption{(Color online.) Momentum differential cross sections for the production of
$X(6900)$ mesons with quantum numbers $J^{PC}=2^{++}$ from the direct processes (1), (2) in the laboratory polar angular range of 0$^{\circ}$--5$^{\circ}$ in the interaction of photons with energy of $E_{\gamma}=$ 35 GeV with $^{12}$C
(left) and $^{184}$W (right) nuclei in the considered theoretical pictures describing their internal structure.}
\label{void}
\end{center}
\end{figure}
\begin{figure}[!h]
\begin{center}
\includegraphics[width=15.0cm]{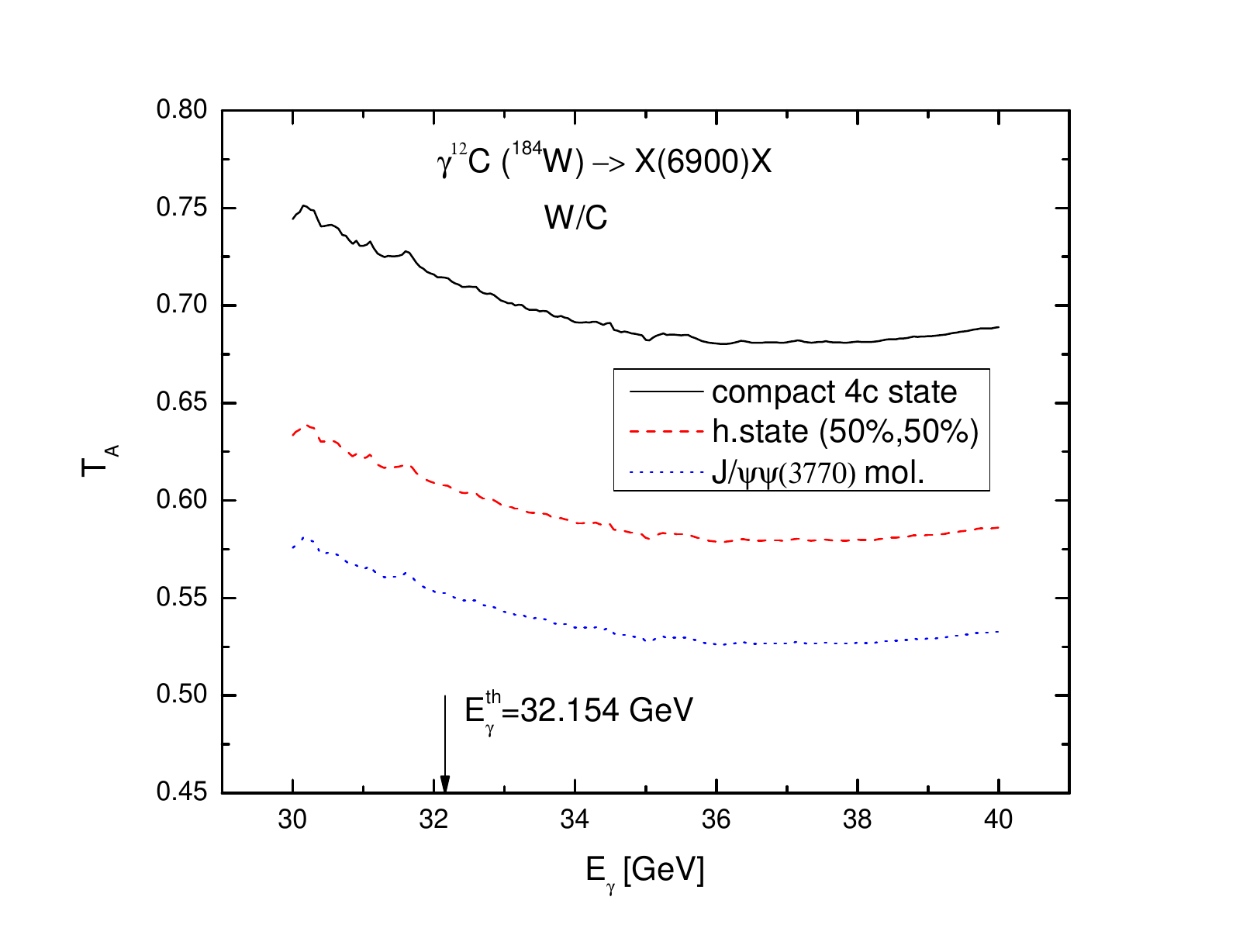}
\vspace*{-2mm} \caption{(Color online.) Transparency ratio $T_A$ for the $X(6900)$ mesons
from the direct processes (1), (2) proceeding on an off-shell target nucleons as a function of the incident photon energy for combination $^{184}$W/$^{12}$C in the considered theoretical pictures describing their internal structure. The arrow indicates the threshold energy for the $X(6900)$ photoproduction on a free target nucleon at rest.}
\label{void}
\end{center}
\end{figure}
\begin{figure}[!h]
\begin{center}
\includegraphics[width=15.0cm]{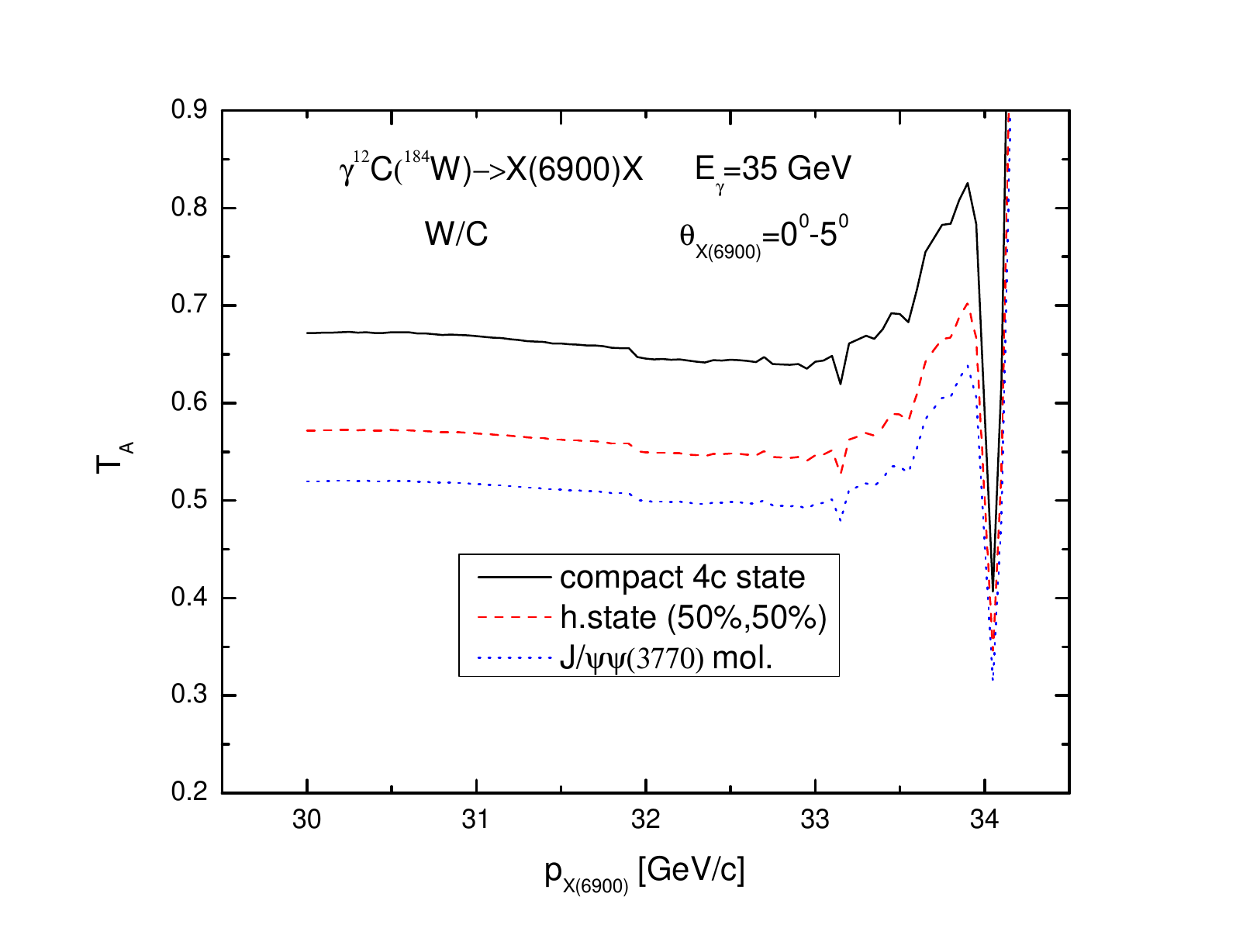}
\vspace*{-2mm} \caption{(Color online.) Transparency ratio $T_A$ for the $X(6900)$ mesons
from the direct processes (1), (2) proceeding on an off-shell target nucleons
as a function of the $X(6900)$ laboratory momentum for incident photon energy of 35 GeV for combination $^{184}$W/$^{12}$C, calculated in the laboratory polar angular range of 0$^{\circ}$--5$^{\circ}$
in the considered theoretical pictures describing their internal structure.}
\label{void}
\end{center}
\end{figure}
\begin{figure}[!h]
\begin{center}
\includegraphics[width=15.0cm]{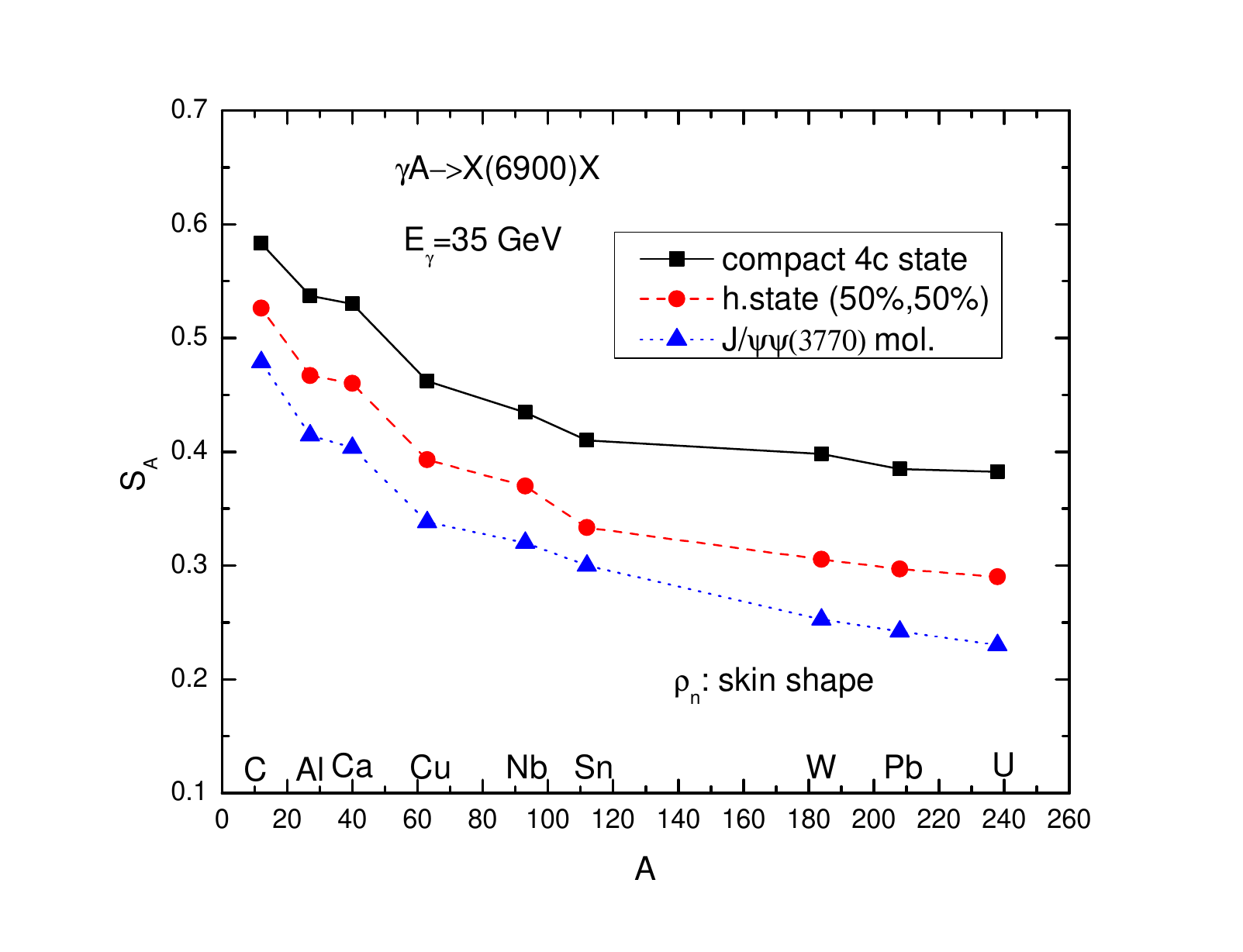}
\vspace*{-2mm} \caption{(Color online.) Transparency ratio $S_A$ for the $X(6900)$ mesons
from the direct processes (1), (2) proceeding on an off-shell target nucleons
at incident photon energy of 35 GeV in the laboratory system as a function of the nuclear mass number $A$,
in the considered theoretical pictures describing their internal structure.
The lines are to guide the eyes.}
\label{void}
\end{center}
\end{figure}
\begin{figure}[!h]
\begin{center}
\includegraphics[width=15.0cm]{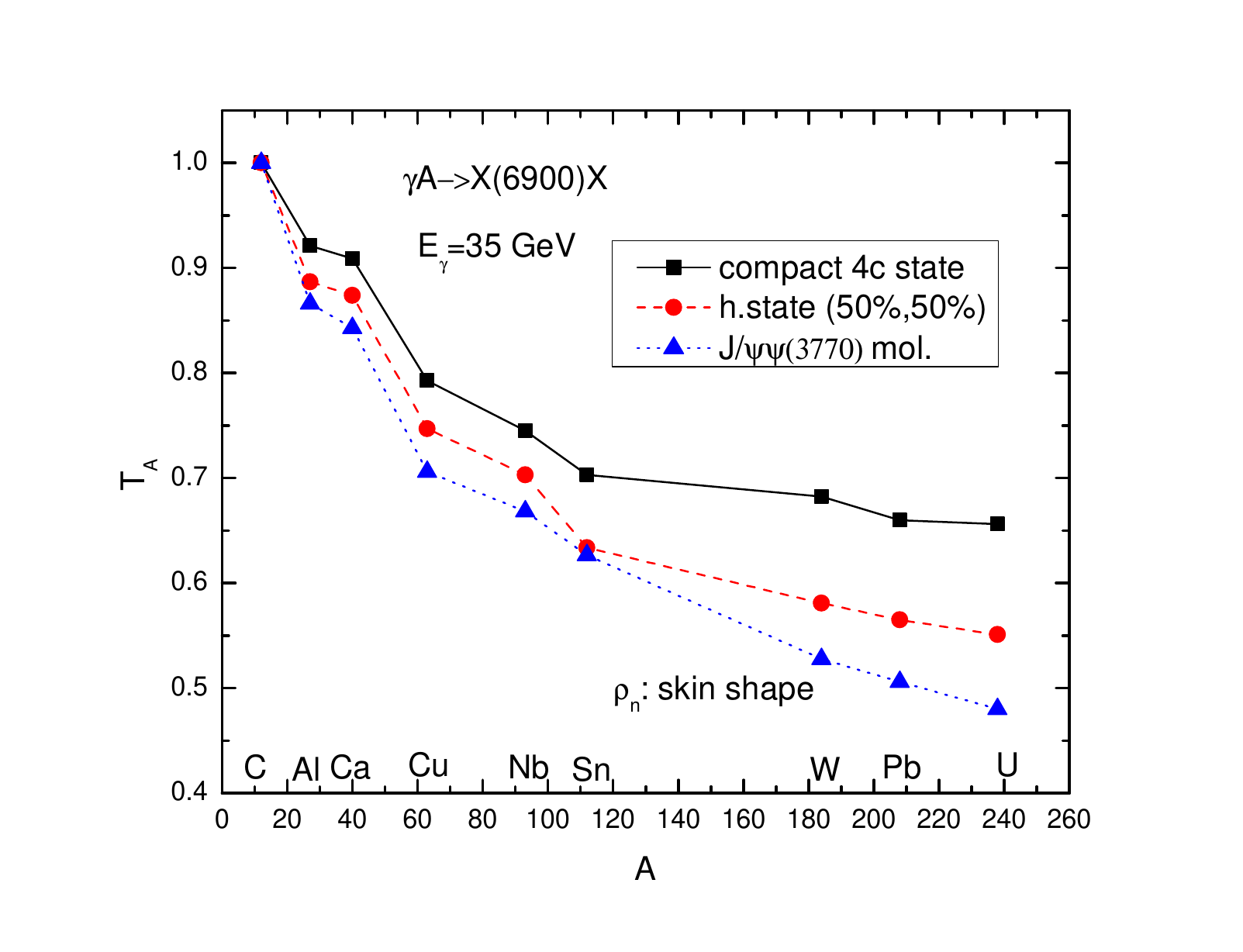}
\vspace*{-2mm} \caption{(Color online.) The same as in Fig. 8, but for the
transparency ratio $T_A$.}
\label{void}
\end{center}
\end{figure}

\section*{3. Numerical results and discussion}

\hspace{1.5cm} The excitation functions for production of $X(6900)$ mesons on $^{12}$C and $^{184}$W target nuclei,
calculated in line with Eq. (3) for three employed options for the $X(6900)$ absorption cross section $\sigma_{X(6900)N}$ as well as for an off-shell target nucleons and for a free ones being at rest with the value of $\sigma_{X(6900)N}=21.2$ mb, are shown in Figs. 3 and 4, respectively. One can see that the influence of the target nucleon Fermi motion on the $X(6900)$ yield is essential only at subthreshold photon energies ($E_{\gamma} < 32.154$ GeV). One can also see yet that for the heavy nucleus $^{184}$W the results depend more strongly on the absorption cross section $\sigma_{X(6900)N}$
and, hence, on the $X(6900)$ internal structure, than those for the light nucleus $^{12}$C. We observe for the $^{184}$W nucleus an experimentally measurable differences $\sim$ 20--30\% between the results corresponding to the calculations with adopted values of this cross section. For the light nucleus $^{12}$C, the same differences as above become smaller. They are $\sim$ 10\% and will probably be experimentally accessible as well in the future high-precision photoproduction experiments at the proposed electron-ion colliders [91, 92] operating in a fixed-target configuration
\footnote{$^)$It is worth noting that the physics opportunities with a fixed-target program at the Electron-Ion Collider EIC, using proton and ion beams on stationary targets, have been discussed in very recent publication [115].}$^)$.
Such measurements look quite optimistic since the $X(6900)$ total photoproduction cross sections have a measurable values $\sim$ 10--40 nb and 100--400 nb for $^{12}$C and $^{184}$W target nuclei, respectively, at above threshold photon energies $\sim$ 33--40 GeV. To inspire the measurements at these colliders also in electroproduction in collider mode, it is desirable to estimate the $X(6900)$ production rates (the event numbers) in the $e^{12}$C and $e^{184}$W reactions
at the $ep$ center-of-mass collision energies $\sqrt{s_{ep}}$ similar to those $\sqrt{s_{{\gamma}p}}$ of the
${\gamma}p$ collision system in a one-year run. For this aim, at first, we translate the $X(6900)$ photoproduction total cross sections, reported above, into the expected total cross sections of the sequences
$e^{12}{\rm C}(^{184}{\rm W}) \to e'X(6900)X$, $X(6900) \to J/\psi{J/\psi}$, ${J/\psi}J/\psi \to 2\mu^+2\mu^-$ by multiplying them on the appropriate branching ratios $Br[X(6900) \to J/\psi{J/\psi}]\approx$ 50\%
\footnote{$^)$The obtained in Refs. [77, 116] within the QCD sum rule method results for the $X(6900)$ total width and for its partial decay width to 2${J/\psi}$ indicate that the branching fraction of its decay to the 2${J/\psi}$ mode is of about 30-45\%. Assuming the decay width of $X(6900) \to 2J/\psi$ to be $\sim$ 60 MeV [60] and using the $X(6900)$ total widths from the LHCb, ATLAS and CMS experiments collected in Table I from [60], we can also get that this branching fraction varies within the range of about 30-75\%. Relying on this, we will adopt in our exploratory study the conservative value of 50\% for it.}$^)$,
$Br[J/\psi \to {\mu^+}{\mu^-}]\approx$ 6\% [17] and by the factor 10$^{-2}$ to account for the fact that approximately two orders of magnitude smaller cross sections are anticipated for the electroproduction compared to the above photoproduction [117]. Then, to estimate the total numbers of the $X(6900)$ events in a one-year run at EicC, one needs to multiply the latter "electroproduction" total cross sections on the carbon and tungsten target nuclei by the integrated luminosity of 60 fb$^{-1}$ [92, 118] as well as by the detection efficiency. With a realistic 50\% detection efficiency [92], we estimate 5.4$\cdot$10$^3$--2.16$\cdot$10$^4$ and 5.4$\cdot$10$^4$--2.16$\cdot$10$^5$ events per year for the $X(6900)$ signal in the cases of the $^{12}$C and $^{184}$W target nuclei, respectively. The respective
sufficiently large event numbers for $X(6900)$ at EIC are 2.7$\cdot$10$^4$--1.08$\cdot$10$^5$ and 2.7$\cdot$10$^5$--1.08$\cdot$10$^6$ with an integrated luminosity of 300 fb$^{-1}$ [91, 118].
We remind that the signal yields of the $X(6900)$ resonance per year reported above were obtained for the poorly known branching fraction of its decay to the 2${J/\psi}$ mode of 50\%. On the other hand, choosing for this fraction the alternative value of $\sim$ 0.5\%, predicted in [119] for the ground state of the $cc{\bar c}{\bar c}$ system with $J^{PC}=2^{++}$, we have that the $X(6900)$ event yields will be two orders of magnitude smaller than those given above and appear to be also considerable yet to be observed at the electron-ion colliders EicC and EIC. Therefore, the excitation function measurements for the exotic $X(6900)$ state both on light and heavy target nuclei also with the $eA$ reactions at these colliders will provide a good platform to deepen our understanding of its internal configuration, once the $X(6900)$ near-threshold proton (and neutron) target photoproduction cross sections will be experimentally known.

For the sake of a clear comparison, we show in Fig. 5 together the absolute momentum distributions of the $X(6900)$ mesons from direct productions processes (1), (2) in the ${\gamma}^{12}$C and ${\gamma}^{184}$W collisions, calculated on the basis of Eq. (15) for three adopted scenarios for their internal configuration for laboratory angles of 0$^{\circ}$--5$^{\circ}$ and for incident photon energy of 35 GeV. We see from this figure that these distributions show a rather sizeable variations, especially for the heavy target nucleus $^{184}$W
\footnote{$^)$Which are similar to those shown in Figs. 3 and 4.}$^)$,
when going from the hadronic molecular to compact tetraquark treatment of the $X(6900)$ state.
Their absolute values have a well measurable strengths $\sim$ 10--20 nb/(GeV/c) and 100--200 nb/(GeV/c) in the central momentum region of 33--33.5 GeV/c in the cases of the $^{12}$C and $^{184}$W target nuclei, respectively.
This behavior of the $X(6900)$ momentum distributions can also be used to decipher its nature from comparison
the present model calculations with data expected from experiments at the electron-ion colliders EicC and EIC.

To provide further guidance for the future experiments and for the sake of illustration, we show in Figs. 6 and 7 the photon energy and the momentum dependences of the transparency ratio $T_A$ for the $X(6900)$ mesons for the $^{184}$W/$^{12}$C combination calculated in line with Eq. (14), using the results presented, respectively,
in Figs. 3, 4 and 5
\footnote{$^)$It should be pointed out that the definition of the transparency ratios $S_A$ and $T_A$ via equations (13) and (14) implies that they should be considered only at above threshold photon energies. But since the right-hand side of Eq. (14) is defined both above and below the respective threshold, we used it in calculating the transparency ratio $T_A$ also at subthreshold photon energies.}$^)$.
It is seen that these dependences can also be used for discriminating between possible scenarios for the $X(6900)$ inner configuration. Furthermore, they depend weakly on the photon energy and on the $X(6900)$ momentum practically at all energies and momenta except of those belonging to the low-energy and to the high-momentum regions. These features can be used as well to better constrain this configuration.

For completeness, we show in Figs. 8 and 9, respectively, the transparency ratios $S_A$ and $T_A$ of the $X(6900)$ mesons in ${\gamma}A$ reactions
($A=$$^{12}$C, $^{27}$Al, $^{40}$Ca, $^{63}$Cu, $^{93}$Nb, $^{112}$Sn, $^{184}$W, $^{208}$Pb, and $^{238}$U)
as functions of the mass number $A$ of the target nucleus.
They have been calculated for the photon energy of 35 GeV in line with Eqs. (13) and (14), correspondingly,
and for three adopted interpretations of the $X(6900)$. It is seen that the ratios $S_A$ and $T_A$ show strong
variations as "functions" of these interpretations and of the mass number. They drop with increasing this number and reach values of the order of 0.2 and 0.5, respectively, for heavy nuclei like $^{208}$Pb and $^{238}$U in the $J/\psi{\psi(3770)}$ molecular scenario -- a large deviation from unity which should be easily seen in a future experiments. On the other hand, there are a sizeable and measurable variations $\sim$ 15, 14\% in the ratio $S_A$
between calculations performed assuming for the $X(6900)$ compact tetraquark and hybrid pictures, hybrid
and pure molecular scenarios already for relatively "light" nuclei like $^{40}$Ca.
For the medium-mass ($^{93}$Nb) and heavy ($^{238}$U) target nuclei these variations are even larger. They are
about 17, 16\% and 32, 26\%, respectively. So, the highest sensitivity of the quantity $S_A$ to the cross section $\sigma_{X(6900)N}$ and, correspondingly, to the $X(6900)$ inner structure is observed for heavy target nuclei.
For the quantity $T_A$ the analogous variations are smaller but yet are experimentally distinguishable in the range of large A. They are about 4, 4\%, 6, 5\% and 19, 15\%, respectively, in the cases of relatively "light", medium-mass and heavy target nuclei. Therefore, we can conclude that the observation of the A-dependences of the transparency ratios $S_A$ and $T_A$, at least, for large mass numbers $A$ in the future photoproduction experiments at electron-ion
colliders would allow to discriminate between compact tetraquark, hybrid and molecular interpretations of exotic $X(6900)$ state and thus to elucidate its nature.

In the end, the absolute and relative observables considered in the present study can be useful to help determine
the $X(6900)$ intrinsic structure.

\section*{4. Epilogue}

\hspace{1.5cm} In the present work the possibility to study the nature of fully-charmed tetraquark state $X(6900)$ from its inclusive photoproduction off nuclei near the kinematic threshold was investigated within the collision model based on the nuclear spectral function. The model accounts for $X(6900)$ production in direct photon--nucleon interactions as well as three different scenarios for its internal structure: compact tetraquark, molecule of the two hidden-charm $J/\psi$ and $\psi(3770)$ mesons and the mixture of both of them. We have calculated within these scenarios the absolute and relative excitation functions on $^{12}$C and $^{184}$W nuclei at near-threshold photon energies of 30--40 GeV, the absolute momentum differential cross sections and ratios of them for the $X(6900)$ meson production off these target nuclei at laboratory polar angles of 0$^{\circ}$--5$^{\circ}$ and for photon energy of 35 GeV as well as the A-dependences of its transparency ratios at photon energy of 35 GeV. We show that the observables considered reveal a definite sensitivity to the $X(6900)$ intrinsic structure. Therefore, they might be useful for the determination of this structure from the comparison of them with the experimental data from the future experiments at the planned electron-ion colliders in the United States and China.
\\

\end{document}